\documentclass[preprint,12pt,3p]{elsarticle}
\usepackage{epsfig,amssymb,amsfonts,bm,color}
\biboptions{sort&compress}

\newcommand{\be}{\begin{equation}}
\newcommand{\ee}{\end{equation}}
\newcommand{\bea}{\begin{eqnarray}}
\newcommand{\eea}{\end{eqnarray}}
\newcommand{\beas}{\begin{eqnarray*}}
\newcommand{\eeas}{\end{eqnarray*}}
\newcommand{\ds}{\displaystyle}
\newcommand{\vev}{{\bm v}}
\newcommand{\Br}{\mbox{Br}}
\renewcommand{\hat}[1]{\not\!{#1}}

\journal{Physics Letters B}

\begin{document}

\begin{frontmatter}

\title{Direct $X(3872)$ production in $e^+e^-$ collisions}

\author[1]{Achim Denig}
\author[2]{Feng-Kun Guo}
\author[3]{Christoph Hanhart}
\author[4,5,6]{Alexey V. Nefediev}

\address[1]{Institute for Nuclear Physics and PRISMA Cluster of Excellence, Johannes Gutenberg University of Mainz, Johann-Joachim-Becher-Weg 45, D-55099 Mainz, Germany}
\address[2]{Helmholtz-Institut f\"ur Strahlen- und Kernphysik and 
Bethe Center for Theoretical Physics, Universit\"{a}t Bonn, D-53115 Bonn, Germany}
\address[3]{Forschungszentrum J\"ulich, Institute for Advanced Simulation, Institut f\"ur Kernphysik and
J\"ulich Center for Hadron Physics, D-52425 J\"ulich, Germany}
\address[4]{Institute for Theoretical and Experimental Physics, B. Cheremushkinskaya 25, 117218 Moscow, Russia}
\address[5]{National Research Nuclear University MEPhI, 115409, Moscow, Russia}
\address[6]{Moscow Institute of Physics and Technology, 141700, Dolgoprudny, Moscow Region, Russia}

\begin{abstract}
Direct production of the charmonium-like state $X(3872)$  in $e^+e^-$ collisions is 
considered in the framework of the vector meson
dominance model. An order-of-magnitude estimate for the width $\Gamma(X\to 
e^+e^-)$ is found to be $\gtrsim$0.03~eV.
The same approach applied to the $\chi_{c1}$ charmonium decay predicts the 
corresponding width of the order
0.1~eV in agreement with earlier  estimates. Experimental perspectives for the 
direct production of the $1^{++}$ charmonia in $e^+e^-$ collisions are briefly discussed.
\end{abstract}


\end{frontmatter}

\section{Introduction}

In 2003 the Belle Collaboration reported the first evidence for the existence 
of a char\-mo\-ni\-um-like state $X(3872)$
\cite{Xobservation}, to be denoted by $X$ for brevity, which possessed 
properties inconsistent with a plain quark--antiquark meson interpretation. 
Later
this state was confirmed independently by many other experimental 
collaborations, see Ref.~\cite{QWGreview} for a
recent review article. The quantum numbers of the $X$ were recently determined by the LHCb 
Collaboration to be $J^{PC}=1^{++}$~\cite{LHCb}.

\begin{figure}[t]
\centerline{
\epsfig{file=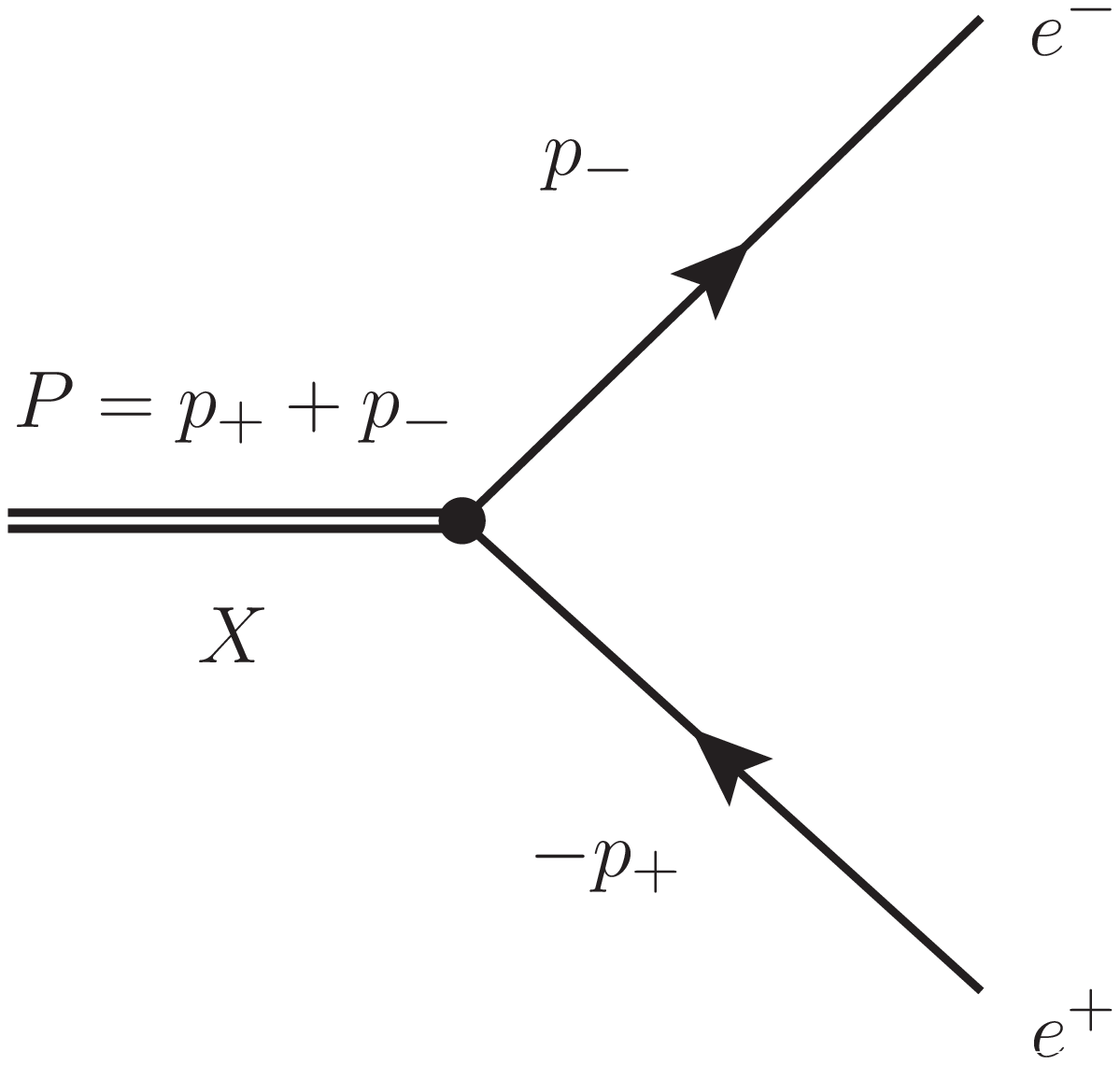,width=0.24\textwidth}
\epsfig{file=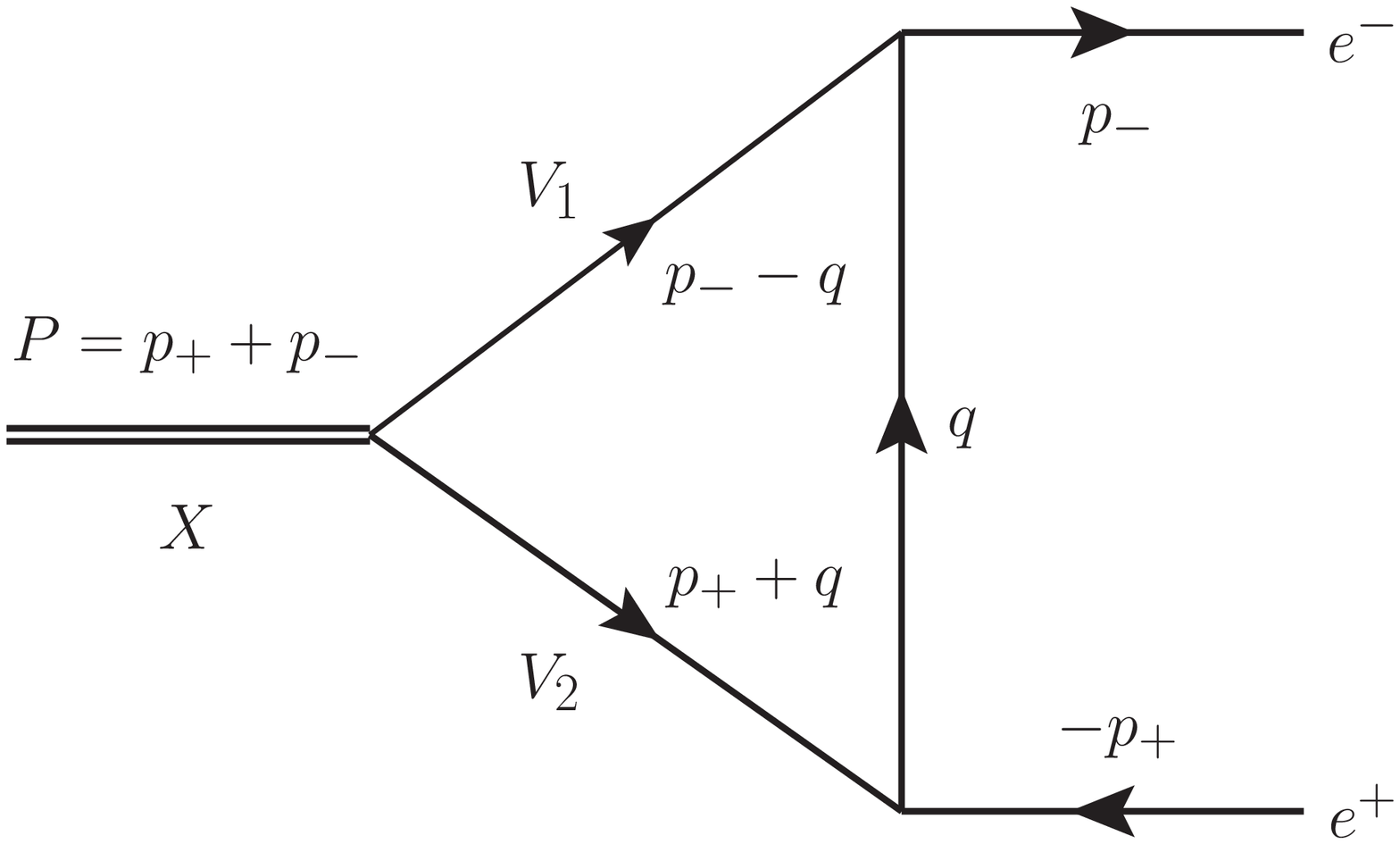,width=0.38\textwidth}
\epsfig{file=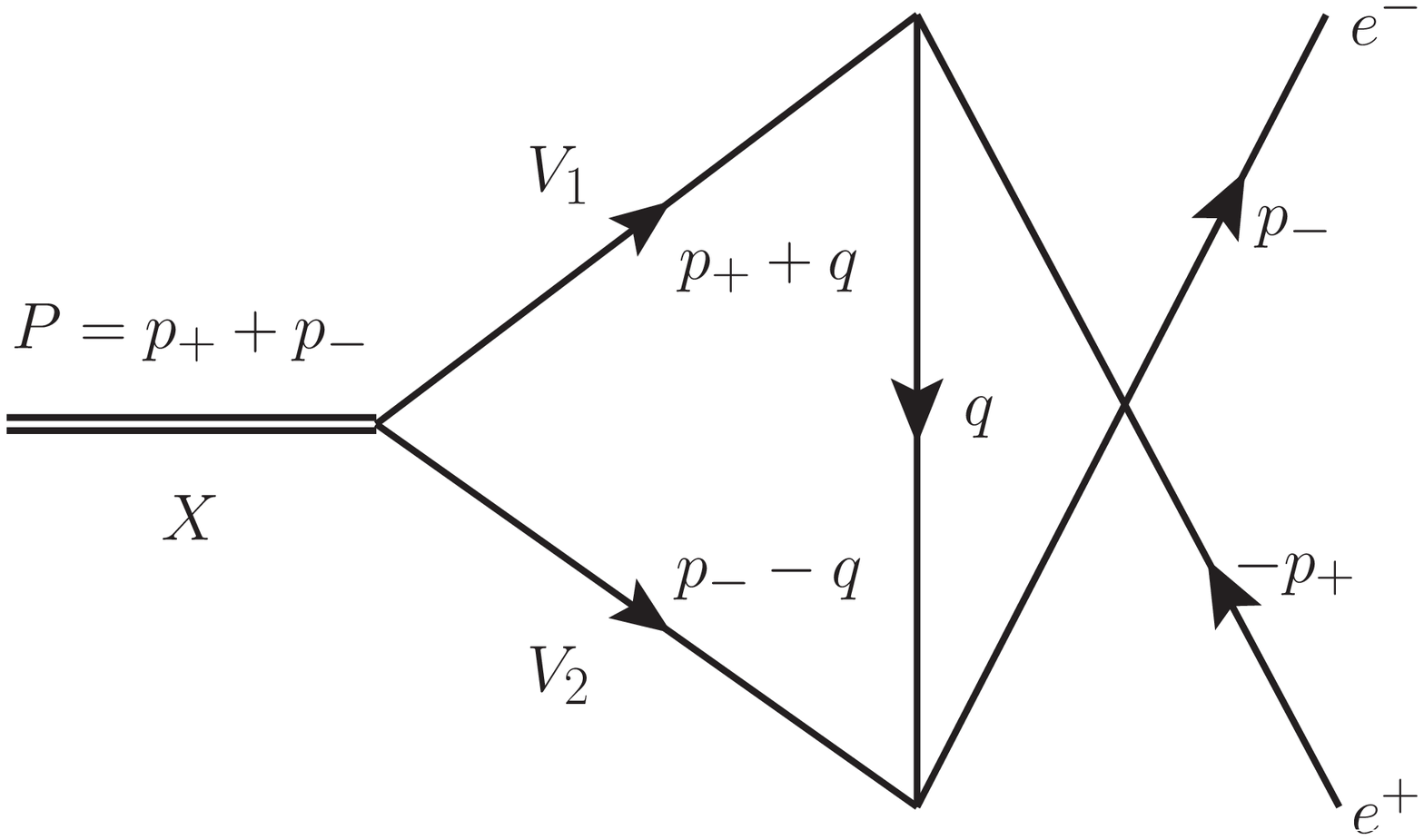,width=0.38\textwidth}}
\caption{Different contributions to the amplitude for the decay $X\to e^+e^-$: the first diagram
accounts for the short-ranged contributions while the other two describe the transitions $X\to
V_1V_2\to e^+e^-$ with $\{V_1,V_2\}$ being $\{\rho,J/\psi\}$, 
$\{\omega,J/\psi\}$, $\{\gamma^*,J/\psi\}$, and
$\{\gamma^*,\psi'\}$.}\label{figXepem0}
\end{figure}

The aim of the present research is to estimate the production rate of the 
$X$ directly in $e^+e^-$ collisions, $e^+e^-\to X$. 
This transition is of course forbidden in $e^+e^-$ annihilation
via a single virtual photon, but can occur via two-photon processes of the kind 
$e^+e^- \to \gamma^* \gamma^* \to X$.
While in the past such a production of a non-vector state was considered as impossible due to
the low production cross section, with the advent of high-luminosity 
accelerators such as BEPC-II, 
operating in the charmonium energy region, a detection might become realistic.

Notice that, while the Landau--Yang theorem forbids
the coupling of an axial-vector state to two real photons, there is no such ban 
for the coupling to two virtual photons. 
To arrive at the desired rate estimate, in this work we parametrise the
vertex $X\to\gamma^*\gamma^*$ in the framework of the vector meson dominance 
(VMD) model,
where  either one of the virtual photons or both are replaced by vector mesons (for details we refer
to Sec.~\ref{xvertex}). In addition, for consistency a short-ranged transition amplitude
needs to be added.
Thus in our model, the decay amplitude is given by the sum of the 
diagrams depicted in Fig.~\ref{figXepem0},
with the vector pairs $\{V_1,V_2\}$ being $\{\rho,J/\psi\}$, 
$\{\omega,J/\psi\}$, $\{\gamma^*,J/\psi\}$, and
$\{\gamma^*,\psi'\}$. Decays of the $X$ into all of these four channels were 
already observed and therefore almost all parameters of the model can be constrained from data.
We stress that for this calculation no specific assumptions need to be
involved for the nature of the $X$ --- the structure information is
encoded in the effective coupling constants. To interpret their values
in terms of different models is a separate issue that goes beyond
the purpose of this work.

\section{Useful experimental information}

The mass of the $X$  is~\cite{PDG}
\be
M_X=(3871.68\pm 0.17)~\mbox{MeV}.
\label{Xmass}
\ee
There exist only upper bounds on its total width~\cite{PDG},
\be
\Gamma_X<1.2~\mbox{MeV},
\label{GX0}
\ee
and on its total production branching fraction in weak $B$-meson 
decays~\cite{total},
\be
\Br(B \to K X) < 3.2 \times 10^{-4}.
\label{babartotal}
\ee
The quantum numbers of the $X$ were determined to be $J^{PC}=1^{++}$~\cite{LHCb}. 

The main
observation modes for the $X$ are the $D^0 \bar D^0 \pi^0$~\cite{DDpi1,DDpi2,DDpi3},
$\pi^+\pi^- J/\psi$ ($\rho J/\psi$) \cite{rho,Belle2pi,LHCb} and $\pi^+\pi^-\pi^0 J/\psi$ ($\omega J/\psi$)
\cite{Babar3pi}, respectively. In addition,
radiative decays $X\to\gamma J/\psi$ and $X\to\gamma\psi'$ (here and in what follows the shorthand notation $\psi'$ is
used for the $\psi(2S)$) were also measured. In particular, the BaBar 
Collaboration reports \cite{Aubert:2008ae} 
\begin{eqnarray}
&&\Br(B^{\pm} \to K^{\pm} X)\Br(X \to \gamma J/\psi)=(2.8 \pm 0.8 \pm 0.2) \times 10^{-6},\nonumber\\[-2mm]
\label{babargamma}\\[-2mm]
&&\Br(B^{\pm} \to K^{\pm} X)\Br(X \to \gamma \psi')=(9.5 \pm 2.9 \pm 0.6) \times 10^{-6},\nonumber
\end{eqnarray}
while the Belle Collaboration gives~\cite{Bhardwaj:2011dj}
\begin{eqnarray}
&&\Br(B^{\pm} \to K^{\pm} X)\Br(X \to \gamma J/\psi)=(1.78^{+0.48}_{-0.44}\pm 0.12) \times
10^{-6},\nonumber\\[-2mm]
\label{bellegamma}\\[-2mm]
&&\Br(B^{\pm} \to K^{\pm} X)\Br(X \to \gamma \psi')<3.45 \times 10^{-6}.\nonumber
\end{eqnarray}
The two results are consistent within errors  for the $\gamma J/\psi$ mode,
however, inconsistent for the  $\gamma\psi'$ mode.
Very recently, the LHCb Collaboration  confirmed that the latter mode has a 
sizable
branching fraction~\cite{Aaij:2014ala},
\be
R_{\gamma\psi}=\frac{\Br(X \to \gamma \psi')}{\Br(X \to \gamma J/\psi)}=2.46\pm 0.64(\mbox{stat})
\pm 0.29(\mbox{syst}). 
\label{LHCbR}
\ee
In order to proceed, we take the averaged value $2.1\times10^{-6}$ 
for the product $\Br(B^{\pm} \to K^{\pm} X)\Br(X \to \gamma J/\psi)$ quoted by 
the Particle Data Group~\cite{PDG} and then use the 
inequality~(\ref{babartotal}) and the LHCb ratio~(\ref{LHCbR}) to arrive at the 
following lower bounds
\be
\mbox{Br}(X\to \gamma J/\psi)> 0.7\,\%,\quad
\mbox{Br}(X\to \gamma \psi')> 1.7\,\%.
\label{branchings2}
\ee
Finally, for our estimates we shall use for the width of the $X$
\be
\Gamma_X=1.0~\mbox{MeV},
\label{Xgam}
\ee
compatible with the upper bound (\ref{GX0}). We also use the following values \cite{PDG} for the masses:
\bea
& m_{\pi^0}=135.0~\mbox{MeV},\quad m_{\pi^\pm}=139.6~\mbox{MeV},\quad m_\rho=775.5~\mbox{MeV},\quad m_\omega
=782.7~\mbox{MeV},&\nonumber\\[-2mm]
\label{masses}\\[-2mm]
&m_{J/\psi}=3096.9~\mbox{MeV},\quad m_{\psi'}=3686.1~\mbox{MeV},\quad M_X=3871.7~\mbox{MeV},&\nonumber
\eea
for the total widths:
\be
\Gamma_\rho=146.2~\mbox{MeV},\quad\Gamma_\omega=8.5~\mbox{MeV}, 
\label{widths1}
\ee
for the partial leptonic widths:
\bea
&\Gamma(\rho\to e^+e^-)=7.0~\mbox{keV},\quad\Gamma(\omega\to e^+e^-)=0.6~\mbox{keV},&\nonumber\\[-2mm]
\label{widths2}\\[-2mm]
&\Gamma(J/\psi\to e^+e^-)=5.6~\mbox{keV},\quad\Gamma(\psi'\to e^+e^-)=2.4~\mbox{keV},&
\eea
and for the branching fractions:
\be
\mbox{Br}(X\to \rho J/\psi)> 2.6\,\%,\quad
\mbox{Br}(X\to \omega J/\psi) > 1.9\,\%.
\label{branchings}
\ee

\section{The $X$-vertex}
\label{xvertex}

According to the diagrams depicted in Fig.~\ref{figXepem0}, the $X$-vertex that feeds the loops
 couples an axial-vector state $X$ to two
vectors $V_1$ and $V_2$. Since the $X$ resides very close to the thresholds of the 
$\rho J/\psi$ and $\omega J/\psi$, the corresponding $X$-vertex can be written 
in a nonrelativistic form,
\be
v_{ijk}(X\to V J/\psi)=\lambda_V\varepsilon_{ijk},\quad V=\rho,\omega,
\label{v}
\ee
where $i$, $j$, and $k$ are contracted with the $X$, $V$, and $J/\psi$ 
polarisation vectors, respectively.

Meanwhile, if one of the vectors is the photon, the nonrelativistic approach 
does not apply\footnote{For a real photon the temporal component 
of the polarisation vector can be set to zero by choosing a suitable gauge. 
Then the $X$-vertex again can be taken in the nonrelativistic form of 
Eq.~(\ref{v}).}.  The
relativistic gauge-invariant $X$-vertex takes the form
\be
v^{\nu\alpha\beta}(X\to\gamma\psi)=\lambda_{\psi}\varepsilon^{\mu\nu\alpha\beta}
k_\mu,\quad
\psi=J/\psi,\psi',
\label{lambdapsi}
\ee
with the Lorentz indices $\nu$, $\alpha$, and $\beta$ being contracted with the 
photon, the $X$, and the $\psi$, respectively, and with
$k^\mu$ denoting the photon 4-momentum.

The coupling constants $\lambda_V$ and $\lambda_\psi$ can be related to the 
corresponding measured partial decay widths of the $X$. In particular, a 
straightforward calculation gives
\be
\Gamma(X\to\gamma \psi)=\Gamma_X\mbox{Br}(X\to\gamma 
\psi)=\frac{\lambda_\psi^2\omega^3}{6\pi M_X^2},\quad
\omega=\frac{M_X^2-m_\psi^2}{2M_X},
\ee
where the experimental branching fractions $\mbox{Br}(X\to\gamma J/\psi)$ and 
$\mbox{Br}(X\to\gamma \psi')$ are
quoted in Eq.~(\ref{branchings2}) and the estimate (\ref{Xgam}) is used for the 
total $X$ width.

The situation with the $\rho J/\psi$ and $\omega J/\psi$ modes is somewhat more 
subtle, since what is actually measured are the branching fractions of the 
processes $X\to\pi^+\pi^- J/\psi$ and $X\to\pi^+\pi^-\pi^0 J/\psi$. We
therefore use
the vertex (\ref{v}) to write the amplitude for the process $X\to V J/\psi\to 
n\pi J/\psi$ ($n=2,3$) in the form
\be
T(X\to n\pi J/\psi)=\lambda_V 
\varepsilon_{ijk}\varepsilon_i(X)\varepsilon_j(J/\psi)G_V(m)v_k(V\to n\pi),
\ee
where $\vev(V\to n\pi)$ is the $V\to n\pi$ vertex, whose explicit form is not 
needed, and
$$
G_V(m)=\frac{1}{m^2-m_V^2+im_V\Gamma_V}.
$$
For the width, one has
\be
\Gamma(X\to n\pi J/\psi)=\frac13\int \sum_{\rm polarisations}
{|T(X\to n\pi J/\psi)|^2}d\tau,
\ee
where for the $X$ at rest as well as for the nonrelativistic $\rho$ or $\omega$ sums over polarisations give
3-dimen\-si\-o\-nal Kronecker deltas. The differential phase space for the final state can be written as
\be
d\tau=d\tau_{n\pi}d\tau_{J/\psi}\frac{dm^2}{2\pi},\quad 
d\tau_{J/\psi}=\frac{p(m)}{4\pi^2 M_X},\quad
p(m)=\frac{1}{2M_X}\lambda^{1/2}(M_X^2,m^2,m_{J/\psi}^2),
\ee
with $d\tau_{n\pi}$ being the phase space for the pions, and 
$\lambda(M^2,m_1^2,m_2^2)$ is the standard triangle function.

Finally, taking into account that
\be
\Gamma(V\to n\pi)=\frac13\int  \sum_{\rm polarisations}|\vev(V\to n\pi)|^2 
d\tau_{n\pi}
\ee
and defining a dimensionless integral over the  mass distribution of the pions
\be
I_V\equiv\int_{nm_\pi}^{m_X-m_{J/\psi}}\Gamma(V\to n\pi)p(m)|G_V(m)|^2m\,dm,
\label{GX}
\ee
one arrives at the relation
\be
\Gamma(X\to n\pi J/\psi)=\Gamma_X\mbox{Br}(X\to n\pi 
J/\psi)=\frac{\lambda_V^2I_V}{2\pi^3M_X},
\ee
which can be used to extract the couplings $\lambda_\rho$ and $\lambda_\omega$ 
with the help of the experimental
branching fractions 
$\mbox{Br}(X\to 2\pi J/\psi)\approx \mbox{Br}(X\to \rho J/\psi)$ and 
$\mbox{Br}(X\to 3\pi J/\psi)\approx \mbox{Br}(X\to
\omega J/\psi)$ quoted in Eq.~(\ref{branchings}).

In Ref.~\cite{2pi3pi} a theoretical analysis was performed of the experimental 
mass distributions for the two-pion and
three-pion final states reported in Refs.~\cite{Belle2pi,Babar3pi}. The results 
of Ref.~\cite{2pi3pi} allow one to
calculate straightforwardly that
\be
I_\rho\approx 0.2,\quad I_\omega\approx 0.02,
\label{Is}
\ee
where the one order of magnitude difference in the two values comes from the 
relatively
 small width of the $\omega$ together with the fact that the nominal $\omega 
J/\psi$ threshold lies
slightly outside of the range of integration in $I_\omega$.

The last missing ingredient is the effective vertex $V\to e^+e^-$ with $V=\rho$, 
$\omega$, $J/\psi$, and $\psi'$, for
which we employ the VMD model. The vector meson--photon vertex respecting gauge 
symmetry can be written as (a detailed discussion of various formulations 
for the vector mesons can be found in Ref.~\cite{Meissner:1987ge})
\be
{\cal L}_{V\gamma}=g_V (\partial^\mu V^\nu-\partial^\nu V^\mu)F_{\mu\nu},
\label{LL}
\ee
where $F_{\mu\nu}$ denotes the usual field
strength tensor for the photon. This leads to a photon--vector meson coupling 
proportional to the 
photon 4-momentum squared, $k^2$. It is this factor that cancels the photon 
propagator in the transition amplitude
$V\to\gamma^*\to e^+e^-$. Therefore the effective $V\to e^+e^-$ coupling 
constant is $2eg_V$, where $g_V$ can be determined from the corresponding leptonic width $\Gamma(V\to 
e^+e^-)$ quoted in Eq.~(\ref{widths2}) with the help of the expression
\be 
\Gamma(V\to e^+e^-)=\frac43\alpha g_V^2m_V,
\label{gV}
\ee
derived straightforwardly from the Lagrangian (\ref{LL}).

\section{Transition amplitude for $X\to e^+e^-$}

In our VMD approach, the total amplitude of the process $X\to e^+e^-$ can be 
written as
\be
T(X\to e^+e^-)=\bar{u}(p_-)V_{\mu}(p_+,p_-)u(-p_+)\varepsilon^\mu(X),
\ee
where $\varepsilon^\mu(X)$ is the $X$ polarisation vector and the full 
$X$-vertex is given by the sum
\be
V_{\mu}(p_+,p_-)=v_\mu^{\rm reg}+v_\mu(X\to \gamma^* J/\psi)+v_\mu(X\to \gamma^*\psi'),
\label{Vvert}
\ee
with $v_\mu^{\rm reg}$ being the regularised contact vertex, and the other two terms are given by the one-loop
amplitudes with $\{V_1,V_2\}=\{\gamma^*, J/\psi\}$, $\{\gamma^*,\psi'\}$. The full transition amplitude is
therefore the sum of the diagrams depicted 
in Fig.~\ref{figXepem0}. Dimensional analysis reveals 
that 
the loop integrals in the amplitudes $v_\mu(X\to\gamma^* J/\psi)$ and $v_\mu(X\to\gamma^*\psi')$ diverge because of the
photon momentum entering the $X$-vertex to preserve gauge invariance, see Eq.~(\ref{lambdapsi}). 
We employ dimensional regularisation with the $\overline{\rm MS}$ subtraction scheme at the 
scale $\mu=M_X$ and absorb the
divergence into the contact vertex $v_\mu^{\rm reg}$. In order to provide a prediction for the rate $X\to e^+e^-$
we need information on the size of this contact term. We here employ two 
different approaches: on one hand, we 
vary the scale $\mu$ in a wide range chosen to be from $M_X/2$ to $2M_X$, which leads to a variation of the
divergent integral of the order of its central value. On the other hand, in order to exclude that the contact term is
enhanced due to
contributions from higher resonances,
we explicitly calculate the transition amplitudes $X\to\rho J/\psi\to e^+e^-$ and $X\to\omega J/\psi\to e^+e^-$, which
contain finite loop integrals only. 

\section{Transition $X\to V J/\psi\to e^+e^-$}

For a given vector meson $V$ ($V=\rho,\omega$), the two one-loop contributions to the amplitude $X\to V J/\psi\to e^+e^-$ 
are shown 
diagrammatically in Fig.~\ref{figXepem0}. The amplitudes read
$$
T_V^{(1)}=2eg_Vg_{J/\psi}\lambda_V\varepsilon_i(X)\varepsilon_{ijk}\int\frac{d^4q}{(2\pi)^4}\bar{u}(p_-)\gamma_j
\hat{q}\gamma_k u(-p_+)G_0(q)G_V(p_--q)G_{J/\psi}(p_++q)
$$
\be
=2eg_Vg_{J/\psi}\lambda_V\varepsilon_i(X)\varepsilon_{ijk}\bar{u}(p_-)\gamma_j\gamma_\mu\gamma_k
u(-p_+)I_{1\mu}(p_+,p_-),
\ee
$$
T_V^{(2)}=2eg_Vg_{J/\psi}\lambda_V\varepsilon_i(X)\varepsilon_{ijk}\int\frac{d^4q}{(2\pi)^4}\bar{u}(p_-)\gamma_k
\hat{q}\gamma_j u(-p_+)G_0(q)G_V(p_++q)G_{J/\psi}(p_+-q)
$$
\be
=-2eg_Vg_{J/\psi}\lambda_V\varepsilon_i(X)\varepsilon_{ijk}\bar{u}(p_-)\gamma_j\gamma_\mu\gamma_k
u(-p_+)I_{2\mu}(p_+,p_-),
\ee
with (the tiny $J/\psi$ width and the electron mass are neglected)
\be
G_0(p)=\frac{1}{p^2+i\epsilon},\quad 
G_{J/\psi}(p)=\frac{1}{p^2-m_{J/\psi}^2+i\epsilon},
\quad G_V(p)=\frac{1}{p^2-m_V^2+im_V\Gamma_V},
\ee 
and
\bea
I_{1\mu}(p_+,p_-)&=&\frac1{i}\int\frac{d^4q}{(2\pi)^4}q_\mu
G_0(q)G_V(p_--q)G_{J/\psi}(p_++q)=\frac{1}{M_X^2}(A_Vk_\mu+B_V
P_\mu),\nonumber\\[-2mm]
\\[-2mm]
I_{2\mu}(p_+,p_-)&=&\frac1{i}\int\frac{d^4q}{(2\pi)^4}q_\mu G_0(q)G_V(p_++q)G_{J/\psi}(p_--q)
=\frac{1}{M_X^2}(A_Vk_\mu-B_V P_\mu),\nonumber
\eea
where $P=p_++p_-$, $k=p_+-p_-$ and the relation $I_{2\mu}(p_+,p_-)=-I_{1\mu}(p_-,p_+)$ was used. Then the full amplitude
reads
\beas
T_V=T_V^{(1)}+T_V^{(2)}&=&\frac{4B_V}{M_X^2}eg_Vg_{J/\psi}\lambda_V\varepsilon_i(X)\varepsilon_{ijk}\bar{u}
(p_-)\gamma_j(\hat{p} _++\hat{p}
_-)\gamma_k u(-p_+)\\
&=&\frac{16B_V}{M_X^2}eg_Vg_{J/\psi}\lambda_V\varepsilon_i(X)
\varepsilon_{ijk}p_k\bar{u}(p_-)\gamma_ju(-p_+),
\eeas
where  the Dirac equation with the electron mass neglected,
$\bar{u}(p_-)\hat{p}_-=\hat{p}_+u(-p_+)=0$, was used. Finally, the width $\Gamma(X\to V J/\psi\to e^+e^-)$
can be evaluated as
\be
\Gamma(X\to V J/\psi\to e^+e^-)=\frac{16|B_V|^2}{3\pi M_X}\alpha g_V^2g_{J/\psi}^2\lambda_V^2,
\label{GX2}
\ee
where the dimensionless coefficient $B_V$ is given by the loop integral,
\be
B_V=\frac1{i}\int\frac{d^4q}{(2\pi)^4}(qP) G_0(q)G_V(p_--q)G_{J/\psi}(p_++q)
=-\frac{1}{32\pi^2}\int_0^1 dx\int_0^{1-x}\frac{(x-y)dy}{a_V^2x+b^2y-xy},
\ee
with
\be
a_V^2=\frac{m_V^2-im_V\Gamma_V}{M_X^2},\quad
b=\frac{m_{J/\psi}^2}{M_X^2}-i\epsilon \ .
\ee
We find from a numerical evaluation
\be
\Gamma(X\to \rho J/\psi\to e^+e^-)\simeq \Gamma(X\to \omega J/\psi\to e^+e^-)\simeq 10^{-7}~\mbox{eV}.
\label{cont}
\ee

The result of Eq.~(\ref{cont}) turns out to be negligible compared to the rate found in the next
section. We therefore regard estimating the contribution of the contact term by varying the integration scale
over a large range as safe.

\section{Transition $X\to \gamma^* \psi \to e^+e^-$}

Similarly to the transition amplitude $T_V(X\to V J/\psi\to e^+e^-)$ studied in 
the previous section, for a given vector
meson $\psi$
($\psi=J/\psi,\psi'$), the two contributions to the amplitude $T_\psi(X\to\gamma^*\psi \to e^+e^-)$ read
$$
T_\psi^{(1)}=\lambda_\psi e g_\psi\varepsilon_\alpha(X)\varepsilon^{\mu\nu\alpha\beta}\int\frac{d^4q}{(2\pi)^4}
\bar{u}(p_-)\gamma_\nu\hat{q}\gamma_\beta u(-p_+)(p_--q)_\mu G_0(q)G_0(p_--q)G_\psi(p_++q)
$$
\be
=\lambda_\psi e g_\psi\varepsilon_\alpha(X)\varepsilon^{\mu\nu\alpha\beta}\bar{u}
(p_-)\gamma_\nu\gamma_\lambda\gamma_\beta u(-p_+)I_{1\mu\lambda}(p_+,p_-),\label{Tpsi1}
\ee
$$
T_\psi^{(2)}=\lambda_\psi e g_\psi\varepsilon_\alpha(X)\varepsilon^{\mu\nu\alpha\beta}\int\frac{d^4q}{(2\pi)^4}
\bar {u}(p_-)\gamma_\beta\hat{q}\gamma_\nu u(-p_+)(p_++q)_\mu G_0(q)G_0(p_++q)G_\psi(p_--q)
$$
\be
=\lambda_\psi e g_\psi\varepsilon_\alpha(X)\varepsilon^{\mu\nu\alpha\beta}\bar{u}
(p_-)\gamma_\nu\gamma_\lambda\gamma_\beta u(-p_+)I_{2\mu\lambda}(p_+,p_-),\label{Tpsi2}
\ee
where
\be
G_0(p)=\frac{1}{p^2+i\epsilon},\quad G_\psi(p)=\frac{1}{p^2-M_X^2a_\psi^2+i\epsilon},\quad
a_\psi^2=\frac{m_\psi^2}{M_X^2},
\ee 
and
\beas
&\ds I_{1\mu\lambda}(p_+,p_-)=\frac1{i}\int\frac{d^4q}{(2\pi)^4}q_\lambda(p_--q)_\mu 
G_0(q)G_0(p_--q)G_\psi(p_++q),&\\[2mm]
&\ds I_{2\mu\lambda}(p_+,p_-)=I_{1\mu\lambda}(p_-,p_+).&
\eeas
After some algebra one finds that 
\be
T_\psi=T_\psi^{(1)}+T_\psi^{(2)}=\lambda_\psi e g_\psi \varepsilon_\alpha\varepsilon^{\mu\nu\alpha\beta}
\bar{u}(p_-)\gamma_\nu\left[I_1\gamma_\mu\gamma_\beta 
+I_2\frac{p_{-\mu}p_{+\beta}}{M_X^2} \right]u(-p_+),
\ee
where the dimensionless integrals $I_1$ and $I_2$ are ($D=4-2\varepsilon$)
\be
I_1=\frac{4i}{D}\int_0^1dx\int_0^{1-x}dy\int\frac{d^Dq}{(2\pi)^D}\frac{q^2}{[q^2-M_X^2x(a_\psi^2-y)]^3},
\ee
\be
I_2=8iM_X^2\int_0^1xdx\int_0^{1-x}(1-2y)dy\int\frac{d^4q}{(2\pi)^4}\frac{1}{[
q^2-M_X^2x(a_\psi^2-y)]^3}.
\ee
A straightforward calculation gives:
\be
I_1^{\rm reg}=\frac{1}{32\pi^2}\left[\ln\frac{M_X^2}{\mu^2}-3+a_\psi^2+\ln a_\psi^2 + (1-a_\psi^2)^2 \bigg(
\ln\left(a_\psi^{-2}-1\right) - i\pi \bigg)
\right]
\ee
and
\be
I_2=\frac{1}{4\pi^2}(1-a_\psi^2)\left[2+(2a_\psi^2-1)\left(\ln\left(a_\psi^{-2}-1
\right)-i\pi\right)\right],
\ee
where, as was explained above, the integral $I_1$ is calculated using the $\overline{\rm MS}$ scheme. The scale
$\mu$ is set equal to $M_X$ for the central value and then varied in the range from $M_X/2$ to $2M_X$ to estimate the
uncertainty.

Finally, the width $\Gamma(X\to \gamma^*\psi \to e^+e^-)$ takes the form
\be
\Gamma(X\to \gamma^*\psi \to e^+e^-)=
\frac{36\pi\alpha I_\psi}{m_\psi\left(1-m_\psi^2/M_X^2\right)^3}\Gamma(X\to\gamma\psi)
\Gamma(\psi\to e^+e^-),
\label{Br0}
\ee
where
\be
I_\psi=48\left[|I_1^{\rm reg}|^2+\frac1{144}|I_2|^2+\frac16{\rm Re}(I_1^{\rm reg}I_2^*)\right],
\quad I_{J/\psi}\approx 3.0\times 10^{-3},\quad I_{\psi'}\approx 2.4\times 10^{-3}.
\ee

Numerical estimates made with the help of Eq.~(\ref{Br0}) give the following lower bounds:
\be
\Gamma(X\to \gamma^* J/\psi \to e^+e^-)\gtrsim 10^{-3}~\mbox{eV},
\label{BrsX1}
\ee
\be
\Gamma(X\to \gamma^* \psi' \to e^+e^-)\gtrsim 0.03~\mbox{eV}.
\label{BrsX2}
\ee
Both rates can only be presented as lower bounds, since for the branching fractions
given in Eq.~(\ref{branchings2})
only the lower bounds exist. Thus, once better data become available, the results of Eqs.~(\ref{BrsX1}) and
(\ref{BrsX2}) may be improved. As discussed before, the contribution of the 
contact term $v_\mu^{\rm reg}$ is estimated by varying the scale $\mu$ in a range as wide as from 
$M_X/2$ to $2M_X$. This leads to a rather conservative estimate for the intrinsic uncertainty of the rates to be of the
order of their central values.

In our approach all parameters are determined from experimental rates. This procedure does not allow
us to extract the signs of the couplings and especially the interference pattern between the amplitude
with the $\gamma J/\psi$ and the amplitude with the $\gamma\psi'$ intermediate state remains undetermined. 
We therefore use Eq.~(\ref{BrsX2}) as the central result and include the possible interference with
the $\gamma J/\psi$ intermediate state as a part of the uncertainty.

It should be stressed that in addition to the uncertainties that arise within the 
formalism used, as 
discussed above, there is also the uncertainty of the model itself.
Unlike effective field theories which have a
controlled uncertainty due to a separation of energy scales and the presence of a 
power counting, our results in Eqs.~(\ref{BrsX1}) and (\ref{BrsX2}) should be 
regarded as an order-of-magnitude estimate, since we are not able to
quantify the intrinsic model dependence.

\section{Discussion}

In this paper we employed a VMD model to estimate the probability of the direct 
production of the charmonium state
$X(3872)$ in $e^+e^-$ collisions, and we arrived at
\be
\Gamma(X \to e^+e^-)\gtrsim 0.03~\mbox{eV}
\label{final}
\ee
which turned out to be dominated by the $\gamma^*\psi'$ intermediate state.
Within our approach the uncertainty of this value can be estimated to be of 
the order of 100\%. This uncertainty contains the one from our ignorance of
a possible short-ranged contribution as well as a possible additional contribution
from the $\gamma^* J/\psi$ intermediate state. Since it is difficult if not impossible
to determine the uncertainty of the model used, we regard the result
of Eq.~(\ref{final}) as no more than a proper order-of-magnitude estimate.

To cross-check
the approach used, one can apply it to the production 
of an ordinary charmonium resonance with the same quantum numbers as the $X$, namely the $\chi_{c1}$.
Within our approach the process
$\chi_{c1}\to e^+e^-$ proceeds predominantly through the $\gamma^* J/\psi$ intermediate state, and its width can be estimated with
the help of an equation similar to Eq.~(\ref{Br0}) with the $X$ replaced by the
$\chi_{c1}$. Using the following $\chi_{c1}$ data~\cite{PDG}:
\be
m_{\chi_{c1}}=3511~\mbox{MeV},\quad \Gamma_{\chi_{c1}}=0.86~\mbox{MeV},\quad
\mbox{Br}(\chi_{c1}\to\gamma J/\psi)\approx 34.8\%,
\label{c1}
\ee
our estimate gives 0.1~eV, and appears to be in a qualitative agreement with 
$\Gamma(\chi_{c1}\to e^+e^-)\simeq 0.46$~eV found in 
Refs.~\cite{kaplan1,kuehn}\footnote{Different approaches were used in 
Ref.~\cite{kaplan1} to calculate the electronic width of the $\chi_{c1}$, and 
the results vary from 0.1 to 0.5~eV. The value 0.46~eV comes from a
VMD model.}, and higher than the lower bound provided by the unitarity limit: 0.044~eV found in 
Ref.~\cite{kaplan1}. 

Experimentally, a production of the $\chi_{c1}$ state in $e^+e^-$ collisions seems very promising
not only due to the high value of $\Gamma(\chi_{c1}\to e^+e^-)$, but also
due to the large branching fraction of $\chi_{c1}$ into $\gamma J/\psi$, which happens to
be a clean experimental signature. Especially, if the $J/\psi$ decay into
$l^+l^-$ ($l=e,\mu$) is considered, detailed studies with the BESIII experiment 
have shown that the only significant background to the $\chi_{c1}$ signal is given by 
the initial state radiation (ISR) production of $l^+l^-$ pairs. 
Neglecting interference effects between the $\chi_{c1}$ and the
ISR amplitudes, the signal to background ratio becomes approximately 10\% if the
value of 0.46~eV is assumed for the electronic width. A discovery of the
reaction $e^+e^- \to \chi_{c1}$ could hence be achieved in an energy scan corresponding to few days of data taking. 

It is instructive to consider in addition the ratio
\be
\Gamma(X\to e^+e^-):\Gamma(\chi_{c1}\to e^+e^-)\gtrsim 1:3,
\label{rat}
\ee
which may cancel some of the uncertainty of the method and thus provides a 
more reliable prediction. 
It turns out that
the most severe suppression factor in the $X$ production as compared to the $\chi_{c1}$ production
comes from the fact that, experimentally, ${\rm Br}(X\to\gamma\psi)\ll
{\rm Br}(\chi_{c1}\to\gamma J/\psi)$ (see Eqs.~(\ref{branchings2}) and (\ref{c1})), while $\Gamma_X\approx
\Gamma_{\chi_{c1}}$. It should be
stressed, however, that the result (\ref{rat}) is based on the upper bound (\ref{babartotal}) on the total $X$
production in the weak $B$-meson decays, so that decreasing this branching would
enhance the width (\ref{final}) and, accordingly, the ratio (\ref{rat}). 
Thus we conclude that the probability of the direct $X$ production in
$e^+e^-$ collisions might  appear in the same ballpark as the probability of the $\chi_{c1}$ production.
\smallskip

The authors are grateful to Ulf-G. Mei\ss{}ner for careful reading of the manuscript and for valuable comments.
This work is supported in part by the DFG and the NSFC through funds provided
to the Sino-German CRC 110 ``Symmetries and the Emergence of Structure in QCD'', 
by CRC 1044 ``The Low-Energy Frontier of the Standard Model'',
by the EU Integrated Infrastructure Initiative HadronPhysics3 (Grant No. 283286), by the Russian presidential programme
for the support of the leading scientific schools (Grant No. NSh-3830.2014.2), and by NSFC (Grant No. 11165005).

\end{document}